# Interaction of Acoustic and Quasi-Elastic Modes in Liquid Water on Nanometer Length Scales

Daisuke Ishikawa[1,2] and Alfred Q.R. Baron[1,2]*

[1]*Materials Dynamics Laboratory, RIKEN SPring-8 Center, Sayo, Hyogo 679-5148, Japan*
[2]*Precision Spectroscopy Division, CSRR, SPring-8/JASRI, Sayo, Hyogo 679-5198, Japan*

We investigate the presence of an acoustic-quasi-elastic interaction contribution in the IXS spectra of liquid water at 301 K using inelastic x-ray scattering with sub-meV energy resolution at momentum transfers $0.77 \leq Q \leq 4.20\,\text{nm}^{-1}$. The contribution appears due to the overlap the acoustic mode with the tail the quasi-elastic mode and is fully consistent with hydrodynamic theory. Incorporating this interaction allows us to describe the dynamic structure factor, $S(Q,\omega)$, without introducing an extra mode, and may help explain earlier contradictory interpretations. The sound velocity, and relative intensity of the quasi-elastic and acoustic mode, plateau for $Q > 2\,\text{nm}^{-1}$ at values consistent with a viscoelastic generalization of the Landau–Placzek relation.

The dynamics of liquids on nm length scales is a fascinating subject for which consensus is still evolving. At long wavelengths, equilibrium liquids are often described by a linearized Navier–Stokes equation (see, e.g., Ref. 1) giving a Rayleigh–Brillouin triplet for the dynamic structure factor, $S(Q,\omega)$, with a quasi-elastic (Rayleigh) mode at zero energy transfer bracketed between the Stokes and anti-Stokes peaks of the acoustic (Brillouin) mode. As the probed length scale shortens into the regime accessible by inelastic neutron or x-ray scattering (INS or IXS), approximately $Q > 0.5\,\text{nm}^{-1}$, one often still observes three peaks in $S(Q,\omega)$, but liquid spectra can blur out, with the acoustic mode width increasing so it sometimes appears as a rather broad feature on the tail of a quasi-elastic peak (Refs. 1–3 and references therein). Further, there is a rich panoply of effects (different types of relaxation, fast sound/positive dispersion, onset of observable transverse dynamics) that potentially impact the behavior at ∼nm length scales. Thus, the meso-scale region of momentum space, where the crossover from continuum to atomistic dynamics occurs in liquids, is both interesting, and challenging.

Interpretations of experimental work on liquid water demonstrate some of these complexities. In particular, measurements of $S(Q,\omega)$ at ∼nm$^{-1}$ momentum transfers have been carried out by INS and IXS (including[4–14]). All the measurements show a large quasi-elastic central peak bracketed by the Stokes and anti-Stokes peaks of a dispersing acoustic mode — a triplet as described above — and the data appears broadly similar across different work. Common features of that work include (1) a quasi-elastic peak that is much bigger than the acoustic mode peaks and (2) the presence of what has been called "fast sound", or "positive dispersion", or "high-frequency sound", as the acoustic mode disperses with a velocity higher than the hydrodynamic (long wavelength) sound velocity. The presence of fast sound is expected from work on other materials (e.g., Ref. 15 is an early example) and also viscoelastic theory or generalized hydrodynamics (e.g., Refs. 1–3, and 16) but the magnitude in water is notably large, with the fast sound velocity roughly double the hydrodynamic value. However, beyond these common features, interpretation of the spectra of water is far from consistent. In particular, as data quality improved, analysis of the data using a simple model (the sum of a Lorentzian and a damped harmonic oscillator, the "L+DHO" model) was augmented by a *second* DHO mode (an "L+2DHO" model) for $S(Q,\omega)$ for $Q \geq 2\text{–}3\,\text{nm}^{-1}$. This extra mode has been suggested to be related to transverse dynamics,[7] or to an anti-crossing behavior[10,12] — while both pictures are discussed in Ref. 17. Meanwhile, analysis of spectra using a memory function approach,[9,13] found no evidence of an additional mode for $2 \leq Q \leq 7\,\text{nm}^{-1}$. Thus, interpretations of the work on liquid water have inconsistencies that have not been reconciled, even over the limited range of momentum transfers ($0.77 \leq Q \leq 4.20\,\text{nm}^{-1}$) that we focus on in the present work.

In fact, the choice of the model used to fit measurements of $S(Q,\omega)$ is important for interpreting most liquid spectra, not only water: acoustic lines for liquids are usually intrinsically broad, and often overlap the quasi-elastic contribution, so that the model choice influences the values determined for parameters, and thus the interpretation of the spectra. While the case of water mentioned above (with at least 3 different interpretations, two with extra modes, 1 without) is extreme, more generally models are used to get information from spectra, to determine mode dispersion, and to suggest the presence, or not, of additional modes (e.g., Refs. 18–20).

Here we demonstrate that interaction between the quasi-elastic mode and the acoustic mode is an important component of the IXS spectra of liquid water for $0.77 \leq Q \leq 4.20\,\text{nm}^{-1}$, and, we suggest, provides a good basis for a common interpretation of the data in this region momentum transfer. Taking this interaction (which appears similar to an extra mode at an energy below the acoustic mode, but with an intensity and position determined by the quasi-elastic and acoustic components) into account, we find the behavior of water at 301 K in this range of momentum transfer can be well described without introducing additional modes. We find the cross-over from hydrodynamic to fast sound occurs at $Q < 2\,\text{nm}^{-1}$ and that the ratio of the intensity of the acoustic mode and the quasi-elastic relaxation is constant for $2 < Q \leq 4.2\,\text{nm}^{-1}$, and consistent with a viscoelastic modification of the Landau–Placzek relation.

The L+DHO or the L+2DHO models used previously may be interpreted in terms of *independent* modes: the quasi-elastic Lorentzian term is typical of relaxation while the DHO contribution(s) describes lifetime broadened (decaying) mode(s).[21] However, naively, there could be interaction between the quasi-elastic and acoustic modes, since the acoustic mode overlaps the tail of the quasi-elastic contribution. One can also make an analogy to Raman scattering where a phonon resonance overlapping a continuum (i.e., non-resonant, energy transfer independent) from electronic scattering can lead to an asymmetric Fano line shape[22] from the interference of the resonant scattering, which changes phase through the transition, and the continuum. In non-resonant IXS, as we use here, the electronic scattering from valence electrons is too weak to observe such interference, however, the tail of the quasi-elastic scattering provides a smooth background that is non-negligible compared to the phonon intensity: one then might expect a similar asymmetric contribution may appear on passing through the acoustic phonon resonance.

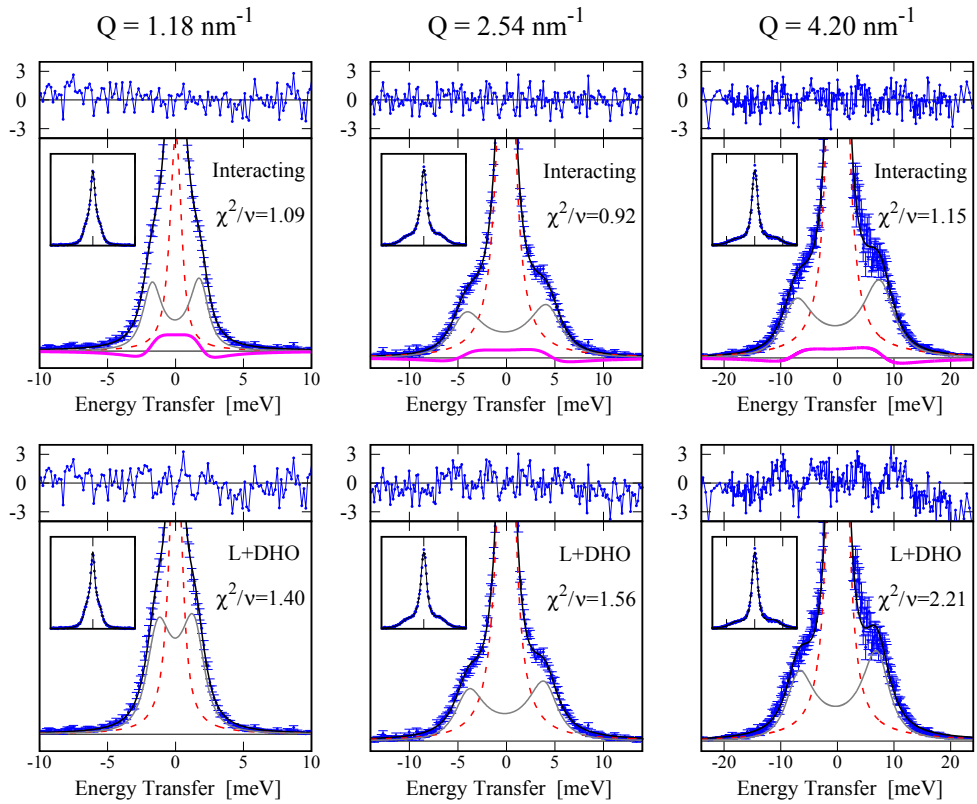

**Fig. 1.** (Color online) Water spectra at $Q = 1.18$, 2.54, and 4.20 nm$^{-1}$ as fit either with a model including interaction (top) or without it (bottom). Insets show the full scale. The lines show the individual components after convolution with the resolution, with the quasi-elastic contribution a dashed line, the DHO solid, and the interaction term (top only) the thick line. The upper panels show the residuals $(y_{\text{data}} - y_{\text{fit}})/\sigma$. The inclusion of the interaction term significantly improves the fit quality without increasing the number of free parameters. The residuals for the L+DHO model also suggest systematic errors, showing correlated oscillations in their deviation from 0, which are removed by using the interacting model.

In fact, if one returns to hydrodynamic theory [e.g., starting from Eq. (5.3.15) of Ref. 1] the dynamic structure factor may be written as

$$S(Q,\omega) = \frac{S(Q)}{\pi}\left[I_0\frac{z_0}{\omega^2+z_0^2} + I_1\frac{2z_1\Omega^2}{(\omega^2-\Omega^2)^2+4z_1^2\omega^2} + I_0 z_0 \frac{\Omega^2-\omega^2}{(\omega^2-\Omega^2)^2+4z_1^2\omega^2}\right], \quad (1)$$

where $\omega$ is the energy transfer and $S(Q)$ is the static structure factor [the integral of $S(Q,\omega)$ over energy transfer]. $I_0, z_0$ are the intensity and width (half width at half maximum, HWHM) of the central Lorentzian mode and $I_1, z_1, \Omega$ are the intensity, width and energy of the sound mode, respectively [normalization requires $I_0 + I_1 = 1$, and relations between the parameters of Eqs. (1) and (5.3.15) of Ref. 1 may be found in the Supplemental Materials[23]]. The first two terms constitute the "L+DHO" model. The last term is the interaction of the acoustic mode with the quasi-elastic mode: its amplitude is proportional to the area and width of the quasi-elastic (Lorentzian) component, while its shape is determined by the position and width of the acoustic (DHO) mode. In hydrodynamics this contribution is effectively included in the "asymmetry parameter", sometimes denoted "b" (e.g., Ref. 24). It integrates to 0, changing sign at the acoustic mode energy, but, to a first approximation (see Fig. 1) may appear similar to a broad extra mode at an energy below that of the acoustic mode. We note Eq. (1) can also be converted to the same form used to describe the rotation-translation coupling in crystals[25] as has been applied to ferroelectric materials,[26] so this type of interaction term is general.

Notably, Eq. (1) is derived in a limit where longitudinal and transverse dynamics are entirely separated, so, a-priori, the interaction is not related to transverse dynamics. Further, the amplitude and shape are entirely determined by the quantities appearing in the Lorentzian and DHO terms, so, when considered as a line-shape for fitting, the interacting form, Eq. (1), has the same number of free parameters as the L+DHO model. Thus, the fact that this expression reproduces water spectra well, as shown below, where the L+DHO model fits poorly, is excellent indication (Ockham's razor) that the interaction is real.

We investigated the dynamic structure factor of water at 301 K at momentum transfers, $0.77 \leq Q \leq 4.20$ nm$^{-1}$, using inelastic x-ray scattering at BL43LXU[27] of the RIKEN SPring-8 center. Runs were made with extremely good energy resolution (down to 0.84 meV, FWHM) using the Si(13 13 13) reflection at 25.7 keV[28] for $0.77 \leq Q < 3.2$ nm$^{-1}$ and using a higher flux setup with the Si(11 11 11) reflection at 21.7 keV for $2.6 \leq Q \leq 4.2$ nm$^{-1}$ (resolution $\geq$ 1.33 meV). Specially designed analyzer masks[29] allowed us to improve count-rates while retaining good momentum resolution. The sample was ultrapure water placed in a cell with two 0.05 mm thick polished single crystal sapphire windows. Additional experimental details can be found in the Supplemental Materials.[23]

The impact of the interaction term can be seen in Fig. 1, where, without increasing the number of fit parameters, there is strong improvement in fit quality, and substantial reduction of correlations in the residuals, relative to the L+DHO model without interaction (see Supplemental Materials for additional information about the fitting[23]). Improvement is visible across the momentum range investigated, with, *always*, the interacting model providing a better fit (see the Supplemental Materials[23]). The improvement in reduced chi-squared by including the interaction is significant: e.g., at 2.54 nm$^{-1}$, the probability that the chi-squared per degree of freedom $\chi^2/\nu \geq$ 1.56 for the L+DHO model is $p < 10^{-5}$ for the $\nu = 128$ degrees in the spectrum ($\chi^2/\nu$ and $p$ are given for several models for all measured spectra in the Supplemental Materials[23]). Of course, one can suggest that the data should be interpreted by adding one or more additional modes to the model: one should pass to an L+2DHO model, or something more complicated, e.g., L+nDHO. However, while this is possible in principle, physical considerations also play a role: not only are the fits better in the interacting model, but that model is directly suggested by theory. Also, as shown in the Supplemental Materials[23] even the L+2DHO, with its full additional line, gives fits that are usually *worse* than the interacting model, and in some cases significantly worse. Better fits, using a model with fewer parameters and a theoretical basis, is strong evidence of validity.

The $Q$-dependence of the fit parameters are shown in Fig. 2. While a detailed analysis of water is beyond the scope of the present work, we discuss several features directly visible in the plots. The sound speed, $\Omega/Q$, and the intensity ratio, $I_0/I_1$, and, to some extent, the quasi-elastic linewidth, $z_0$, saturate at constant values for $Q \geq 2$ nm$^{-1}$. The magnitude of the fast or high-frequency sound, 2.91 km/s from Fig. 2(a), is about double that of the hydrodynamic value, 1.50 km/s at this temperature. This is reasonably consistent with, or slightly lower than, other work — though other work uses different fitting models. We note that the plateau in the sound speed, intensity ratio and quasi-elastic width suggest that if one considers the water response in the context of a viscoelastic transition starting at small $Q$, then that transition has largely completed by $Q \sim 2$ nm$^{-1}$. Meanwhile, an approximately linear $Q$-dependence of the acoustic mode width, $z_1$, is possible within viscoelastic models over a limited range of momentum transfers, though it represents an interplay of different quantities.[16]

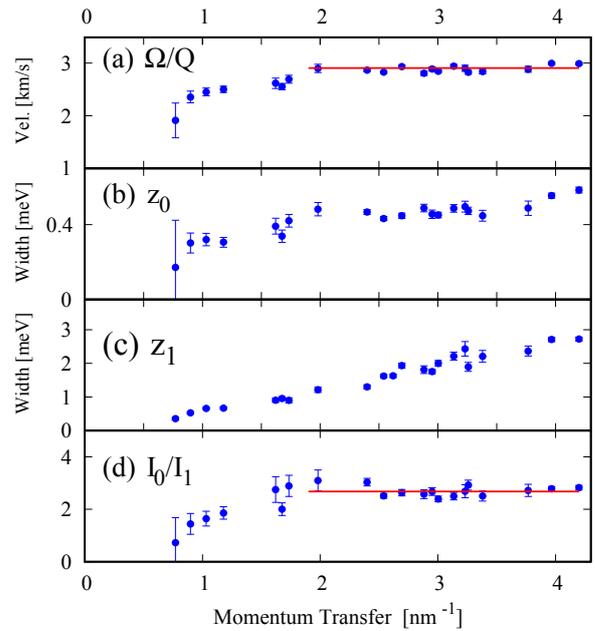

**Fig. 2.** (Color online) $Q$-dependence of parameters from fits to the interacting model — see text for discussion. The lines are constant fits to the indicated regions, and are in good agreement with a viscoelastic modification of the Landau–Placzek relation.

We briefly focus on the intensity ratio, $I_0/I_1$, as, viewed from a long-wavelength limit, this is perhaps the most surprising characteristic of the water spectra. In particular, light scattering measurements at $Q \sim 0.02$ nm$^{-1}$ show a very small quasi-elastic intensity, typically a few % of the acoustic mode intensity.[30,31] This is in good agreement with the Landau–Placzek relation, $I_0/I_1 = \gamma - 1$, where $\gamma = c_p/c_\nu$ is the specific heat ratio: $\gamma = 1.03$ for water at room temperature so one expects, and observes, $I_0/I_1 \approx 0.03$.[30,31] In contrast, at $Q \geq 2$ nm we observe roughly a 100-fold increase to $I_0/I_1 \approx 2.67$ [the line in Fig. 2(d)]. However, this change is consistent with a viscoelastic generalization of the Landau–Placzek relation — something, that, to the best of our knowledge has not previously been mentioned in the context of spectra measured in the present range of momentum transfer. In particular, when one is above the crossover to fast sound, a generalization of the Landau–Placzek relation is[16]

$$\frac{I_0}{I_1} \approx \gamma \left(\frac{c_\infty}{c_0}\right)^2 - 1. \quad (2)$$

Taking $\gamma \approx 1.03$ for water at room temperature near atmospheric pressure, $c_0 = 1.50$ km/s[32] and $c_\infty = 2.91(2)$ km/s directly from Fig. 2(a), Eq. (2) predicts $I_0/I_1 \approx 2.88(4)$ which is in acceptable agreement with the 2.67(5) of the fit in Fig. 2(d). This suggests that the relaxation represented by the quasi-elastic mode in the spectra may be predominantly that responsible for the transition to the higher sound speed.

We now revisit some of the previous work measuring $S(Q,\omega)$ for water in this range of momentum transfer. Very early work using both neutrons and x-rays analyzed the data in terms of a L+DHO model (or, sometimes, 3 Lorentzians), however, more recent work, with better quality data introduced a second DHO mode, passing to an L+2DHO model, where the second mode was attributed variously to the

 

onset of observable transverse dynamics or to the presence of an anti-crossing.[7,10,12,17] Meanwhile, more recently, some workers have taken the interpretation in terms of a transverse mode as being accepted (e.g., Ref. 14). We suggest that, at least over the range of momentum transfers investigated here, the extra mode may have been an artifact of using a non-interacting L+DHO model as a starting point for the description. In fact, our observed lack of any extra mode in the investigated $Q$ range is consistent with a memory function approach to the analysis of water spectra where no extra mode was observed for $2 \leq Q \leq 7\,\text{nm}^{-1}$.[9,13] This agreement is not surprising given that the memory function effectively used in that work has been shown[33] to be equivalent to the hydrodynamic line-shape. However, as has been our experience, and noted by others, analysis using memory functions is "usually affected by strong correlations among fitting parameters"[34] so the direct frequency domain approach used here, and its interpretation, may offer advantages in practical treatment of $S(Q,\omega)$. We also note, that above $3.9\,\text{nm}^{-1}$ we do observe a slight increase in the quasi-elastic linewidth which may hint at a need to consider an additional component at higher $Q$, and that component might be related to transverse dynamics.

In sum, we have shown that including interaction between the quasi-elastic mode and the acoustic mode in water spectra, $S(Q,\omega)$, at momentum transfers of 0.77 to $4.20\,\text{nm}^{-1}$ gives good fits to the measured data without introducing extra modes, simplifying the interpretation of the data. The interaction is fully consistent with hydrodynamic theory from a linearized Navier–Stokes equation.[1] We suggest that Eq. (1) is then a good, accurate and straightforward-to-interpret, form for spectra at low momentum transfers. This will be useful for fitting and understanding liquid spectra, and results of molecular dynamics simulations, and as a starting point for investigating more complex effects. In the present work we also show that the main cross-over from hydrodynamic long-wavelength behavior to the high frequency behavior in ambient water occurs at small momentum transfers, $Q < 2\,\text{nm}^{-1}$, and that a viscoelastic modification of the Landau–Placzek relation accounting for fast sound reasonably gives the intensity ratio of the quasi-elastic and acoustic parts of the response, suggesting the quasi-elastic mode observed in IXS in this range may be predominantly due to the relaxation causing fast sound.

**Acknowledgements** IXS measurements were made at BL43LXU of the RIKEN SPring-8 Center. A.B. is grateful to Hajime Tanaka for a comment at a workshop reminding him of Mountain's work and to Sergey Vakhrushev for recalling the paper by Stock et al.

**Author contributions** Experimental work was done by D.I. and A.B. The approach to data analysis via Eq. (1) was suggested by A.B. and the data analysis and fitting was carried out by A.B. with input from D.I. The paper was written by A.B. with input from D.I.

*baron@spring8.or.jp

Supplemental Materials for

# Interaction of Acoustic and Quasi-Elastic Modes in Liquid Water on Nanometer Length Scales

Published as a Letter in the Journal of the Physical Society of Japan (2021)

Daisuke Ishikawa and Alfred Q.R. Baron*

*Materials Dynamics Laboratory, RIKEN SPring-8 Center, 1-1-1 Kouto, Sayo, Hyogo 679-5148 JAPAN*

*Precision Spectroscopy Division, CSRR, SPring-8/JASRI, 1-1-1 Kouto, Sayo, Hyogo 679-5198 JAPAN*

**Section S1: Experimental Conditions and Data Treatment**

For IXS measurements the water cell was mounted in a vacuum chamber with a cleanup pinhole (0.3 mm diameter) at 25.7 keV (Si (13 13 13)) and a cleanup slit (0.3 x 2mm$^2$) at 21.7 keV (Si (11 11 11)), placed just upstream of the cell (in vacuum) to reduce backgrounds. The samples used were 5 mm thick at 25.7 keV, and 2 mm at 21.7 keV. With the pinhole in place, and a Soller slit on the outgoing beam path [1], empty cell backgrounds were always much smaller than the signal, and in fact were negligible for higher Q. The temperature at the sample was not actively controlled but the hutch temperature, as monitored near the sample position, was measured to be an average of 27.4 C with a diurnal variation of < 0.5C from the average. Momentum transfers are expected to be accurate to better than 0.02 nm$^{-1}$ and the momentum resolution was flat topped with a width <20% of the momentum transfer, and typically closer to 10%.

In order to compare our measured result to the various models of $S(Q,\omega)$ we took the classical model function for $S(Q,\omega)$ given by eqn. 1, scaled it by a detailed balance factor ( $DB(\omega) = x/(1-e^{-x})$, $x = \hbar\omega/k_B T$ ) and convolved the result with the energy resolution for the relevant analyzer/mask combination. That resolution was determined in the conditions used in the experiment by measuring scattering from plexiglass near its structure factor maximum and then removing the small phonon contribution to the plexiglass as described in [2]. As the masks were tailored to provide good resolution, momentum acceptances were near top-hat distributions (see [1]). For completeness, fits were done both with and without averaging $S(Q,\omega)$ over the relevant range of momentum transfers, but this had small impact on the results – and the data presented includes the averaging. The parameters of the model function were optimized by minimizing chi-squared with respect to the (background subtracted, when necessary) data using code based on the MINUIT package [3] first using the MIGRAD routine and then MINOS to find the final minimum. Throughout this paper, error bars represent 1-sigma deviations on counting statistics propagated, when needed, according to standard methods.

---

* baron@spring8.or.jp

**Section S2: Goodness of Fit**

Figure S1 shows the goodness of fit parameter, $\chi^2/\nu$, where $\nu$ is the number of degrees of freedom, for the best fits for several models. The interacting model uniformly gives a significantly improved value of this goodness of fit parameter compared to an L+DHO model, and, in most cases gives a comparable or better fit than even the L+2DHO model, with sometimes, the L+2DHO being significantly worse. This can be emphasized by plotting the probability, p, of have $\chi^2/\nu$ larger than the observed value (see, e.g., discussion in [4]) as may be calculated using an incomplete gamma function [5]: the interacting model usually gives fits with p>10%, while the L+DHO model often has p<<1% and even occasionally p<10$^{-4}$. The L+2DHO, which doubles the number of phonon lines available, so is a major increase in fitting freedom compared to the interacting or L+DHO models, does better than the L+DHO model, but usually not as well as the single-phonon line interacting model (eqn. 1), and, occasionally, the L+2HO is in fact significantly worse than the interacting model.

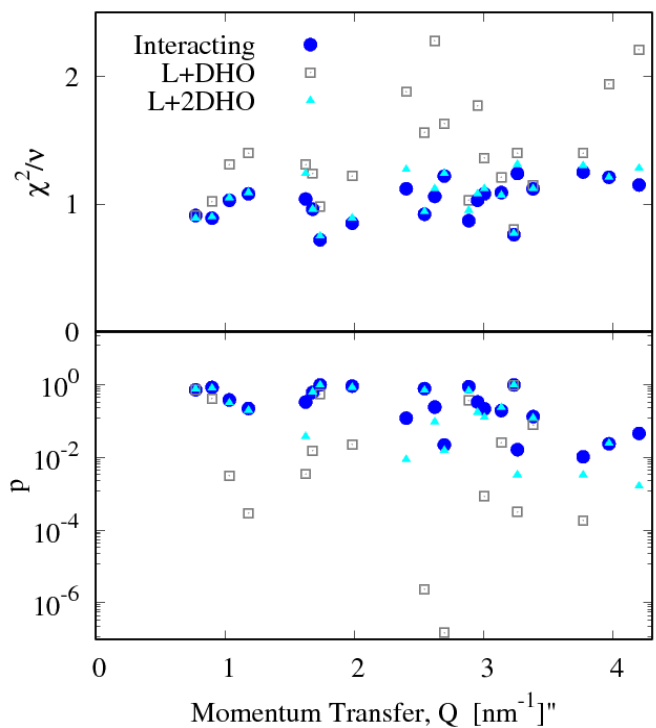

Figure S1. Goodness of fit for several models. The top plot shows the chi-squared, $\chi^2$, per degree of freedom, while the lower plot shows the probability of getting that value or larger for a data set assuming the measured data points have normally distributed random errors. See discussion in text. Note that some p values for the L+DHO fits are below 10$^{-7}$ and do not appear in the lower plot.

**Section S3: Acoustic Mode Energy:** $\Omega$, $\omega_s = \sqrt{\Omega^2 - z_1^2}$, $\omega_{pk}$

There have been different choices made as to what parameter is used to discuss the acoustic mode energy in liquids, with some work choosing $\Omega$ (including much of the referenced work on water), other work focusing on $\omega_s = \sqrt{\Omega^2 - z_1^2}$, and still other work choosing the peak, $\omega_{pk}$, of the longitudinal current-current correlation function, $J(Q,\omega) = \omega^2 S(Q,\omega)/Q^2$. The last is convenient as it can be determined directly from the shape of $S(Q,\omega)$, and, in principle, is independent of the choice of the model for $S(Q,\omega)$. We prefer $\Omega$ because (1) in the limit where the quasi-elastic mode is weak or narrow ($I_0 \ll 1$ or $z_0 \ll \Omega$) then $\omega_{pk} \approx \Omega$ and (2) if the quasi-elastic mode is strong or broad, or if one does make more complicated models with possible additional modes, it seems reasonable to discuss the acoustic mode energy as a parameter that is *not* affected by tails of the other modes. For completeness, we note that for the interacting model, one has

$$\omega_{pk,\text{interacting}} = \Omega \left[ 1 - \frac{1}{2} \frac{I_0}{I_1} \frac{z_0 z_1}{\Omega^2} + \frac{5}{8} \left( \frac{I_0}{I_1} \frac{z_0 z_1}{\Omega^2} \right)^2 + H.O. \right] \qquad (SM1)$$

and we plot the various mode energies in figure S2. As suggested, $\omega_{pk} \approx \Omega$.

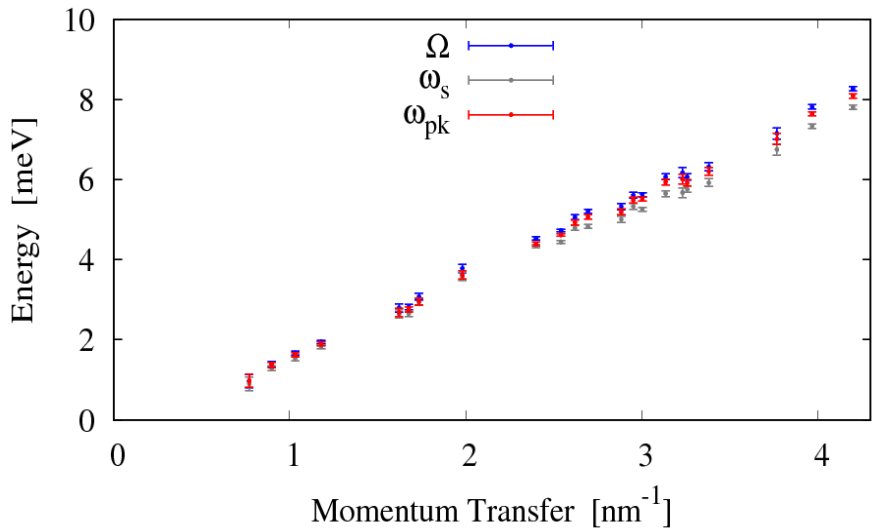

Figure S2. Different possible choices for acoustic mode energies as discussed in the text.

**Section S4: Conversion of Parameters**

For convenience and completeness we give relations between the parameters of equation 5.3.15 of [6] and equation 1 of the main text:

$$z_0 = \chi k^2 \qquad (SM2a)$$

$$z_1 = \Gamma k^2 = \tfrac{1}{2}\left[v_l + (\gamma - 1)\chi\right]k^2 = \tfrac{1}{2}v_l k^2 + \tfrac{1}{2}(I_0/I_1)z_0 \qquad (SM2b)$$

$$v_l k^2 = 2z_1 - (\gamma - 1)z_0 = 2z_1 - z_0 I_0/I_1 \qquad (SM2c)$$

$$\gamma = 1 + I_0/I_1 \qquad (SM2d)$$

$$c^2 k^2 = \omega_s^2 = \Omega^2 - z_1^2 \qquad (SM2e)$$

Where k is used in [6] instead of $Q$, $\chi$ is the thermal diffusivity, $v_l$ is the longitudinal viscosity and $\Gamma$ is the acoustic mode width. It is perhaps worth noting that while equation 1 can be negative if $z_1$ is small compared to $z_0$ and $I_1$ is small compared to $I_0$, the parametrization above prevents this, as, when $I_0$ becomes large, $z_1$ becomes large compared to $z_0$ (eqn. SM2b): the hydrodynamic formulation does not allow for an acoustic mode that is simultaneously weak and narrow compared to the central line, as is consistent with all spectra observed in our experience.

**References for these Supplemental Materials**
1) A.Q.R. Baron, D. Ishikawa, H. Fukui, and Y. Nakajima, AIP Conf. Proc. **2054**, 20002 (2019).
2) D. Ishikawa and A.Q.R. Baron, J. Synch. Rad. **28**, 804 (2021).
3) F. James and M. Roos, Comput. Phys. Commun. **10**, 343 (1975).
4) P.R. Bevington and D.K. Robinson, *Data Reduction and Error Analysis for the Physical Sciences*, Second Edi (WCB McGraw-Hill, Boston, 1992).
5) A. Gil, J. Segura, and N.M. Temme, SIAM J. Sci. Comput. **34**, A2965 (2012).
6) J.P. Boon and S. Yip, *Molecular Hydrodynamics* (Dover Publications, Mineola, New York, 1980).